\documentclass[aps,prd,preprint,tightenlines,nofootinbib,showpacs,fixfloat]{revtex4}
\usepackage{epsfig}
\usepackage{graphicx}
\usepackage{dcolumn}
\usepackage{bm}
\usepackage{overpic}
\usepackage{subfigure}
\usepackage{float}

\def \bea{\begin{eqnarray}}
\def \beq{\begin{equation}}

\def \eea{\end{eqnarray}}
\def \eeq{\end{equation}}
\def \ege1{E(\gamma_{\rm E1})}

\def \gme1{\gamma_{\rm E1}}

\newcommand{\epem}{e^+e^-}
\newcommand{\pipihc}{\pi^+\pi^-h_c}
\newcommand{\eepipihc}{\epem\to\pipihc}

\newcommand{\etahc}{\eta h_c}
\newcommand{\eeetahc}{\epem\to\etahc}

\begin{document}


\title{A Review of $h_c(^1P_1)$, $\eta_c(1S)$ and $\eta_c(2S)$}

\author{\small
Jianming~Bian
\vspace{0.2cm}\\
\vspace{0.2cm} {\it
School of Physics and Astronomy\\
University of Minnesota, Minneapolis, MN 55455\\
$~$\\
$~$\\
Presented at the 5th International Workshop on Charm Physics\\
Honolulu, Hawai'i ,  May 14$-$17, 2012\\
}}

\vspace{0.4cm}

\begin{abstract}

Recent experimental results on charmonium $h_c(^1P_1)$, $\eta_c(1S)$ and $\eta_c(2S)$ from Belle, BaBar, CLEO and BESIII are reviewed. $h_c$ production and properties, the $\eta_c(1S)$ lineshape and the observation of $\eta_c(2S)$ in $\psi'$ decays are discussed.

\end{abstract}

\pacs{14.40.Gx, 13.25.Gv, 13.20.Gd, 12.38.Qk}
\maketitle
\section{\bf Introduction}
\bigskip

The charmonium family is a great laboratory for precision tests of the quark model,  because of their relative immunity from complications like relativistic
effects and the large value of the strong coupling constant $\alpha_s$. In this talk, a brief review of recent experimental results on charmonium $h_c(^1P_1)$, $\eta_c(1S)$ and $\eta_c(2S)$ is presented. Although these states were predicted just after the discovery of $J/\psi$, their properties were not very clear for a long period. $h_c$ is the mostly recently discovered charmonium states; recent studies uncovered its production and properties. $\eta_c(1S)$ is the lowest-lying $S$-wave spin-singlet charmonium state and has been observed through various processes. Its ``inconsistent'' lineshapes in different production modes inspired a couple of precise measurements in the last few years. The $\eta_c(2S)$ is the first radial excitation of the $\eta_c$ charmonium ground state. After 30 years of searching, it was recently observed in charmonium transitions, having been observed in $B$ decays and $\gamma\gamma$ fusion previously. Results in this talk come from BESIII, CLEO, BaBar and Belle. BESIII and CLEO study charmonium from $\psi'$ decays and $e^+e^-$ annihilation near $D\bar{D}$ threshold. They provide very clean and simple environments. BaBar and Belle are $B-$factories, which produce charmonium via $\gamma\gamma$ fusion and $B$ decays. The advantage of studying charmonium in a $B-$factory is the relatively large statistics and reconstruction efficiency.

\section{\bf $h_c$}
\bigskip

Of the charmonium states below $D\bar{D}$ threshold, the $h_c(1^{1}P_{1})$ is experimentally the least accessible. That is because it cannot be produced directly in $e^+e^-$ annihilation, or appear in the electric dipole transition process of a $J^{PC} = 1^{--}$ charmonium state. Statistics and photon detection also made it very challenging for early experiments to observe $h_c$ in charmonium transitions. 

Information about the spin-dependent interaction of heavy quarks can be obtained from precise measurement of the $1P$ hyperfine mass splitting $\Delta~M_{hf}\equiv\langle M(1^3P)\rangle- M(1^1P_1)$,
where $\langle M(1^3P_{J})\rangle=(M(\chi_{c0})+3M(\chi_{c1})+5M(\chi_{c2}))/9=
3525.30\pm0.04$~MeV/$c^2$~\cite{ref:PDG_2012} is the spin-weighted centroid of
the $^3P_J$ mass and $M(1^1P_1)$ is the mass of the singlet state
$h_c$. A non-zero hyperfine splitting may give indication of non-vanishing
spin-spin interactions in charmonium potential models~\cite{swanson}.

The first evidence of the $h_c$ state was reported by the Fermilab
E760 experiment~\cite{ref:E760hc} and was based on the process
$p\bar{p}\to \pi^0J/\psi$.  This result was subsequently excluded by the successor
experiment E835~\cite{ref:E835hc}, which investigated the same reaction
with a larger data sample.   E835 also studied the process
$p\bar{p}\to h_c\to\gamma\eta_c(1S)$, in this case finding an $h_c$ signal. Soon after
this the CLEO collaboration observed the $h_c$ and measured its mass~\cite{ref:cleohc05,ref:cleohc08} by studying the decay chain $\psi(3686)\to\pi^0 h_c, h_c\to\gamma\eta_c(1S)$ in $e^+e^-$ collisions.  CLEO subsequently presented evidence for $h_c$ decays to multi-pion final states~\cite{ref:cleohc09}. Since these data were collected in 2009, BESIII has put lots of effort into measuring the properties of $h_c$.

To study the decay $\psi'\to\pi^0 h_c, h_c\to\gamma\eta_c(1S)$, three methods have been used:

\begin{itemize}
\item Inclusive: In the inclusive mode, only the $\pi^0$ is detected and the $h_c$ are recognized as a peak in the $\pi^0$ recoil mass spectrum. The $\pi^0$ momentum in $\psi'\to\pi^0 h_c$ is about 85 MeV. The inclusive yield is used to extract the absolute branching ratio of $\psi'\to\pi^0 h_c$. This mode has the largest background.

\item E1-tagged: Detecting the $\pi^0$ and the E1 transition $\gamma$ from the $h_c\to\gamma\eta_c(1S)$ (~500MeV). The E1-tagged signal yield is proportional to the product branching ratio $B(\psi'\to\pi^0 h_c)\times B(h_c\to\gamma\eta_c)$. Combining with the inclusive measurement, E1-tagging also provides the absolute branching ratio of $h_c\to\gamma\eta_c(1S)$. The background for this mode is smaller than that for the inclusive mode.

\item Exclusive: Reconstructing $\pi^0$, E1$\gamma$ and all of the decay products of the $\eta_c(1S)$ in $\psi'\to\pi^0 h_c, h_c\to\gamma\eta_c(1S)$. Here all final-state particles are detected and energy-momentum conserving kinematic fits can be used to improve the resolution. This method has small background and provides the best $h_c$ mass and width measurement. The yield is proportional to $B(\psi'\to\pi^0 h_c)\times B(h_c\to\gamma\eta_c)\times B(\eta_c\to X_i)$, where the $X_i$ refer to specific final states in the $\eta_c(1S)$ decay.

\end{itemize}

\begin{figure}[htbp]
\begin{center}
  \includegraphics[width=8cm]{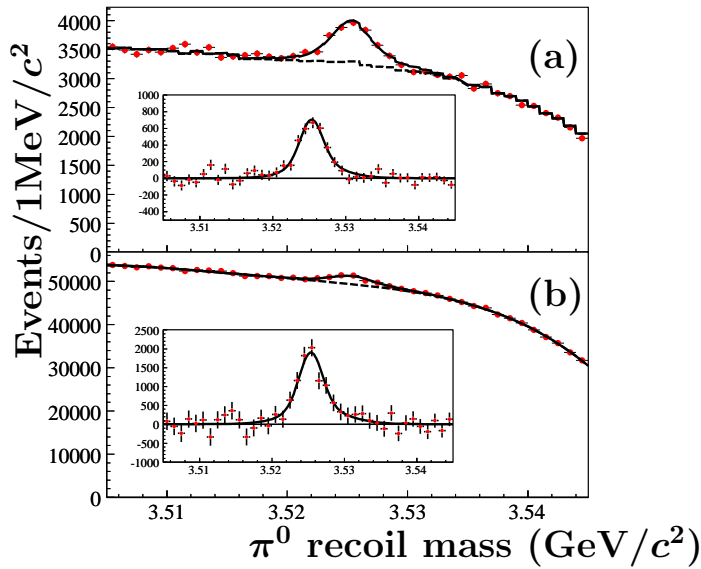} \put(-40,180){\bf
    \large~(a)} \put(-40,90){\bf
    \large~(b)}\put(-230,70){\rotatebox{90}{\bf \boldmath \large
      Events/1MeV/$c^2$}}\put(-140,-5){\bf\boldmath \large $\pi^{0}$
    recoil mass~(GeV/$c^2$)}
\caption{Inclusive/E1-tagged $h_c$ measurement from BESIII: (a)~The $\pi^0$ recoil mass spectrum and fit for the
  $E1$-tagged analysis of $\psi'\to \pi^0h_c,
  h_c\to\gamma\eta_c(1S)$;~(b)~the $\pi^0$ recoil mass spectrum and fit for
  the inclusive analysis of $\psi'\to\pi^{0}h_c$. Fits are shown as
  solid lines, background as dashed lines. The insets show the 
  background-subtracted spectra.}\label{hc-FitData}
\end{center}
\end{figure}

The observation of $h_c$ from CLEO used the E1-tagged and exclusive modes. Using inclusive and E1-tagged modes, BESIII first measured the absolute branching ratios $B(\psi(3686)\to\pi^{0}h_c)=(8.4\pm1.3(stat.)\pm1.0(syst.))\times10^{-4}$ and $B(h_c\to\gamma\eta_c)=(54.3\pm6.7\pm5.2)\%$~\cite{ref:bes3hc10}. These results are consistent with theoretical expectations and make it possible to extract absolute $h_c$ cross sections/branching ratios~\cite{KYT}$-$\cite{ref:hctheorygod02}. In the same paper, BESIII also determined the mass and width of $h_c$ to be $M(h_c)=3525.40\pm0.13\pm0.18$~MeV and $\Gamma(h_c)=0.73\pm 0.40\pm 0.50$~MeV ($90\%$ confidence level upper limit is 1.44MeV), which agree with CLEO's observation. With this mass value, the $P-$wave hyperfine splitting is $-0.10\pm 0.22$ MeV, consistent with zero. Fig.~\ref{hc-FitData} shows the $h_c$ signals and fits in the inclusive and E1-tagged $\pi^0$ recoil mass spectrum in $\psi'$ decays~\cite{ref:bes3hc10}. 

\begin{figure}[tbhp]
\begin{center}
\epsfig{file=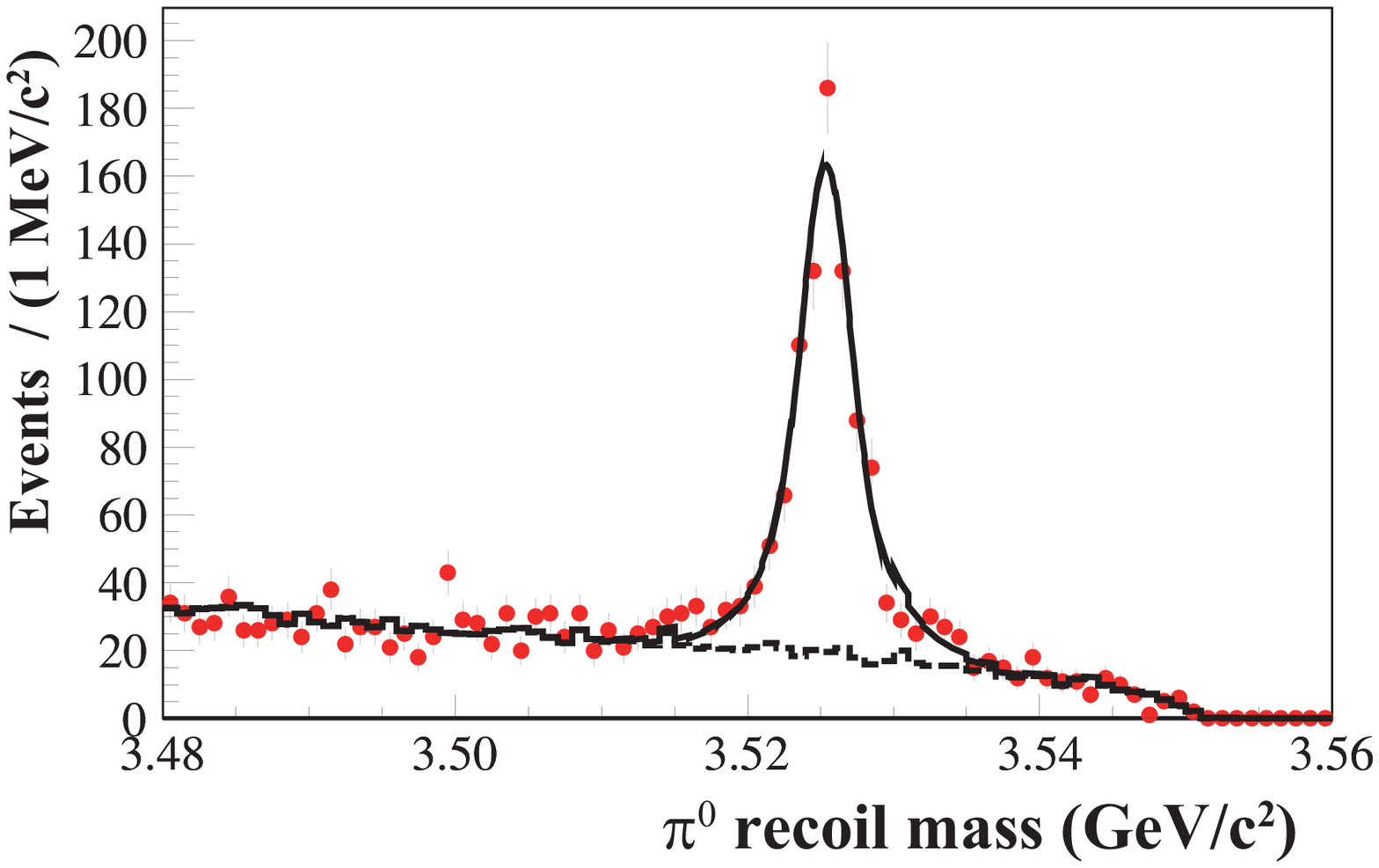,width=14cm} \caption{The $\pi^0$
recoil mass spectrum in $\psi(3686)\to\pi^0{h}_c, {h}_c\to\gamma\eta_c(1S)$,
$\eta_c \to X_i$ summed over the 16 final states $X_i$ in BESIII's $h_c$ exclusive study.  The dots with error bars represent the $\pi^0$ recoil mass spectrum in data, The solid line shows the total
fit function and the dashed line is the background component of the fit.
\label{fig:fithcpi0tot}}
\end{center}
\end{figure}

A detailed  study of 16 exclusive channels is in progress at BESIII~\cite{Olsen:2012xn}.  The aim is to obtain the most precise $h_c$ resonance parameters and study $\eta_c(1S)$ line-shape parameters in the $E1$ transition $h_c\to\gamma\eta_c(1S)$. Preliminary results of this study are $M(h_c)=3525.31\pm0.11\pm0.15$ and $\Gamma(h_c)=0.70\pm0.28\pm0.25$~MeV. The summed fit results are shown in Fig.~\ref{fig:fithcpi0tot}.

\begin{figure}[htbp]
\begin{center}
\epsfig{file=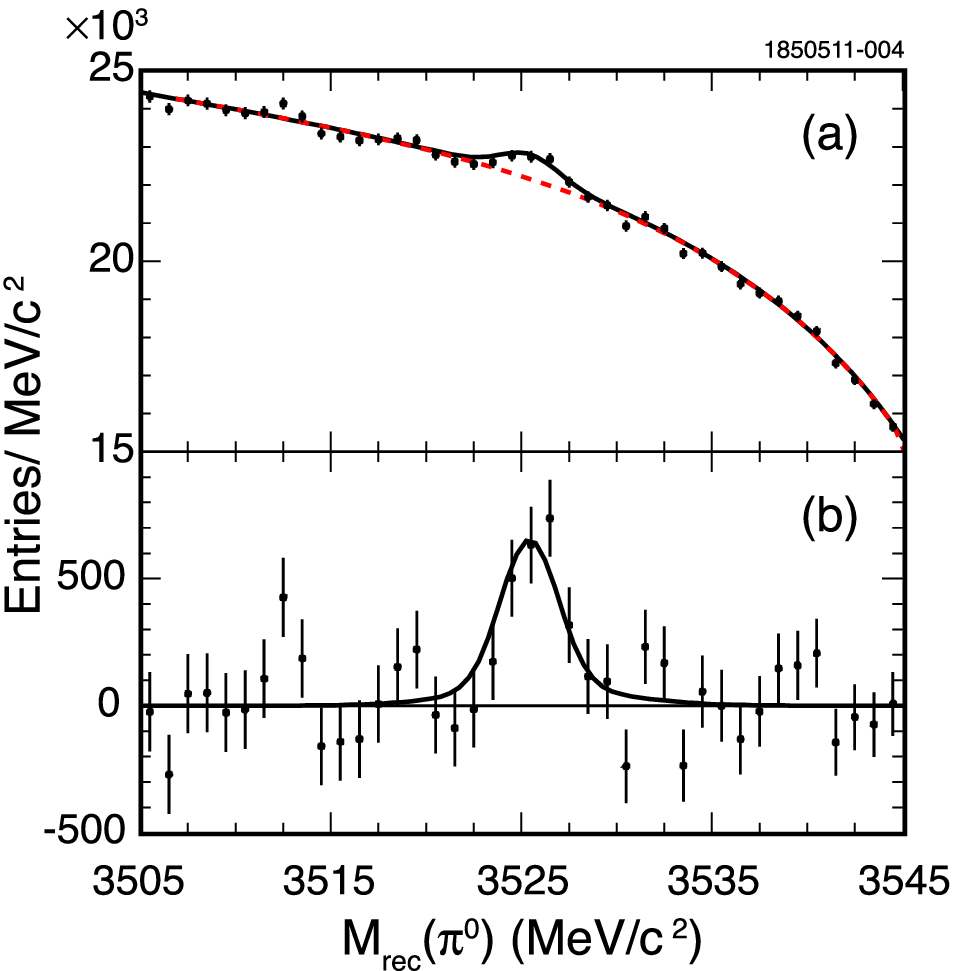, width = 8 cm}
\caption{Inclusive/E1-tagged $h_c$ measurement from CLEO: (a) Fit to the inclusive $\pi^0$ recoil mass spectrum of $\psi'$.
(b) As in (a) but with the background fit from (a) subtracted.
\label{fig:fig4}}
\end{center}
\end{figure}

CLEO has confirmed BESIII's inclusive measurement result~\cite{Ge:2011kq}. By rejecting very asymmetric $\pi^0\to\gamma\gamma$ decays, CLEO also observed $h_c$ signal in the inclusive $\pi^0$ spectrum in $\psi'$ decays (Figure \ref{fig:fig4}). $B(\psi'\to\pi^0 h_c)$ is determined to be $(9.0\pm1.5\pm1.3)\times10^{-4}$, which is consistent with the BESIII measurement.

\begin{figure}[htbp]
\begin{center}
\epsfig{file=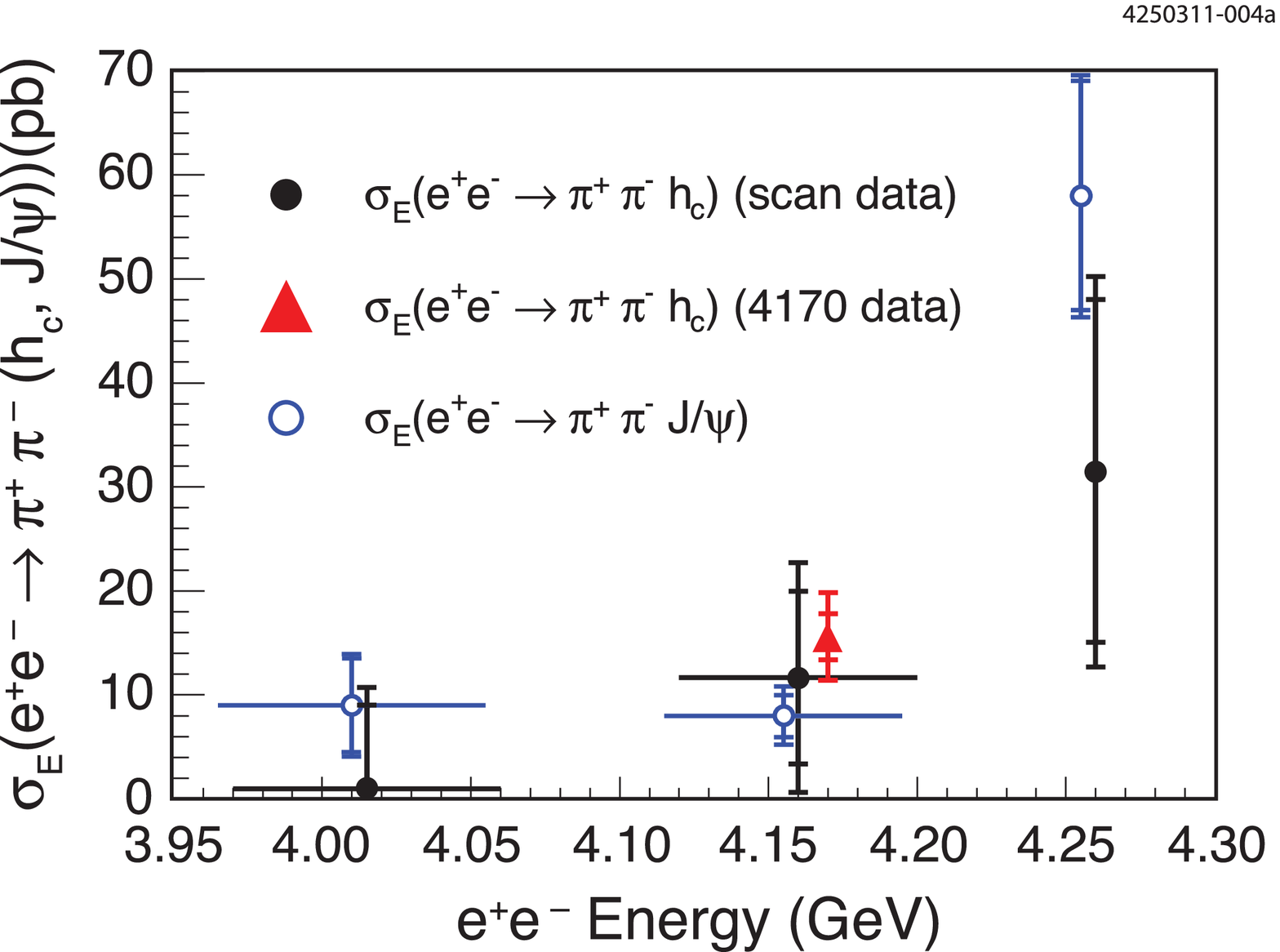, width = 8 cm}
\caption{Cross sections of $\eepipihc$ as a function of center-of-mass energy measured by CLEO.  The triangle shows the cross section for $\eepipihc$ at $E_{cm}=4170$~MeV; the closed circles are for the same process at other center-of-mass energies.  For reference, the $\epem\to\pi^+\pi^-J/\psi$ cross section is indicated by open circles.  The inner error bars are the statistical errors; the outer error bars are the quadratic sum of the statistical and systematic errors.
\label{fig:SCAN}}
\end{center}
\end{figure}

Beyond $\psi'\to\pi^0 h_c$, CLEO has made an important discovery of $h_c$ production in $e^+e^-\to\pi^+\pi^- h_c$ at $\sqrt{s} =4170$\,MeV using 586 pb$^{-1}$ of $e^+e^-$ annihilation data. 10$\sigma$ signal for $h_c$ was found in the decay
$e^+e^-(4170)\to\pi^+\pi^- h_c, h_c\to \gamma \eta_c, \eta_c\to$ 12 decay modes.  This result demonstrats a new prolific source of $h_c$ and has inspired the Belle collaboration to search for $ h_b(1P, 2P)$ in $e^+e^-$ annihilations at    = 10.685 GeV using the same technique. CLEO also finds evidence for $\eeetahc(1P)$ at $4170~$MeV at the $3\sigma$ level, and sees hints of a rise in the $\eepipihc(1P)$ cross section at $4260~$MeV~\cite{CLEO:2011aa}. The $\pipihc$ cross sections measured by CLEO at different center-of-mass energies are summarized in Fig.~\ref{fig:SCAN}.

\section{\bf $\eta_c(1S)$}
$\eta_c(1S)$ is the lowest-lying $S$-wave spin-singlet charmonium state. Although it has
been known for about thirty years~\cite{Himel:1980dj}, its resonance parameters are still interesting. 

For a long period, the measurements of the $\eta_c(1S)$ width from $B-$factories
and from charmonium transitions were inconsistent. In PDG10~\cite{ref:PDG_2010}, confidence level of the global fit  is only 0.0018 for the $\eta_c(1S)$ mass and only 0.0001 for the $\eta_c(1S)$ width. These discrepancies can be attributed to poor statistics and inadequate consideration of interference between $\eta_c(1S)$ decays and non-resonant backgrounds. The experimental confusion introduced difficulties in the determination of the charmonium $1S$ mass hyperfine $M(J/\psi)-M(\eta_c)$. With old $\eta_c(1S)$ parameters in PDG10, $1S$ hyperfine mass splitting is $116.6\pm1.2$ MeV, away from theoretical predictions~\cite{etachyperfine}.

Recent studies by Belle, BaBar, CLEO, and BESIII~\cite{Vinokurova:2011dy,delAmoSanchez:2011bt,Mitchell:2008aa,BESIII:2011ab}, with large data samples and careful consideration of interference, obtained similar $\eta_c(1S)$ width and mass results in two-photon-fusion production and $\psi'$ decays.

\begin{figure}[htbp]
\begin{center}
\epsfig{file=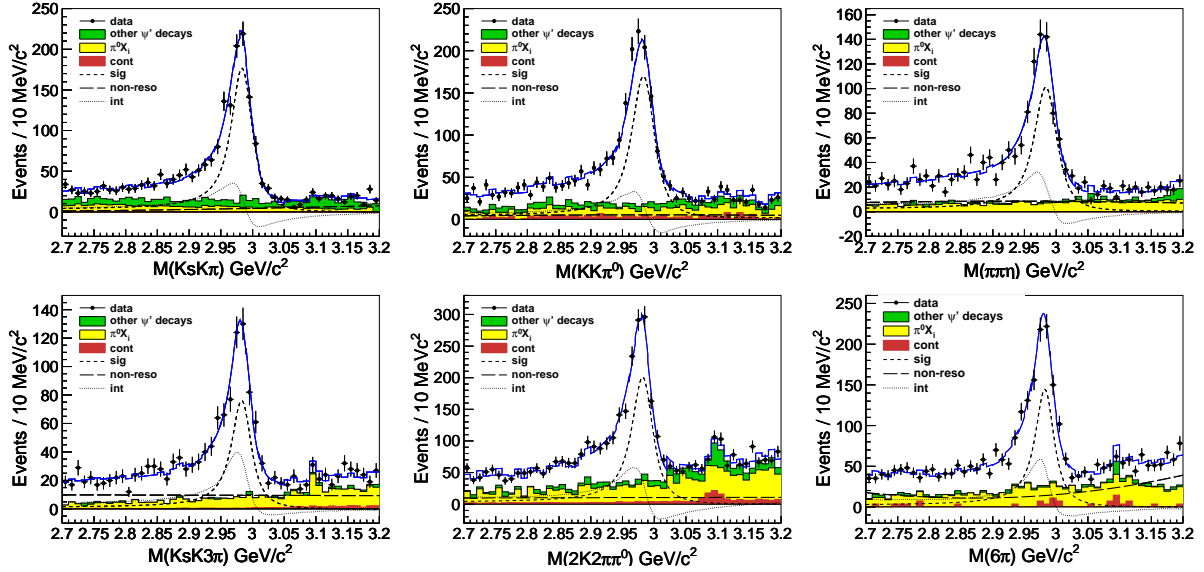, width = 16 cm}
  \caption{$\eta_c(1S)$ lineshape measurement in $\psi'\to\gamma\eta_c(1S)$ from BESIII: The $M(X_i)$ invariant mass distributions for the decays 
  $K_S K^+\pi^-$, $K^+K^-\pi^0$, $\pi^+\pi^-\eta$, $K_S K^+\pi^-\pi^+\pi^-$, $K^+K^-\pi^+\pi^-\pi^0$ and $\pi^+\pi^-\pi^+\pi^-\pi^+\pi^-$, respectively, with the fit results (for  the constructive solution) superimposed. Points are data and the various curves are
  the total fit results. Signals are shown as short-dashed lines, the non-resonant
  components as long-dashed lines, and the interference between them as dotted lines.
  Shaded histograms are in red/yellow/green for continuum/$\pi^0 X_i$/other $\psi'$ decays backgrounds.  The continuum backgrounds for $K_S K^+\pi^-$ and $\pi^+\pi^-\eta$ decays are negligible.}
  \label{fig:metac}
\end{center}
\end{figure}

In 2009, CLEO observed a distortion in the $\eta_c(1S)$ line shape in $\psi'\to\gamma\eta_c(1S)$. CLEO  concluded that the distortion is caused by photon-energy dependence of the magnetic dipole transition rate (hindered-M1 transition)~\cite{Mitchell:2008aa}. This observation inspired BESIII's $\eta_c(1S)$ line shape study via $\psi'\to\gamma\eta_c(1S)$ with a 106M-event $\psi'$ sample~\cite{BESIII:2011ab}. In BESIII's $\eta_c(1S)$ analysis, 
$\eta_c(1S)$ is reconstructed with six decay modes: $K_S K^+\pi^-$, $K^+K^-\pi^0$, $\pi^+\pi^-\eta$, $K_SK^+\pi^-\pi^+\pi^-$, $K^+K^-\pi^+\pi^-\pi^0$, and $3(\pi^+\pi^-)$.  A simultaneous fit to these channels is performed. The $\eta_c(1S)$ Breit-Wigner is weighted by  an $E_\gamma^7$ factor to account for the energy dependence of the hindered-$M1$ transition. Interference with background from non-resonant $\psi'$ decays is also considered. The new BESIII mass and width values are $M(\eta_c) = 2984.3\pm 0.6\pm 0.6~\rm{MeV}/c^2$ and $\Gamma(\eta_c) = 32.0\pm 1.2\pm 1.0~ \rm{MeV}$. They agree well with results from $B-$factories. Using only the new BESIII $\eta_c(1S)$ mass value, the $J/\psi-\eta_c(1S)$ hyperfine mass splitting is $112.6\pm0.8$ MeV, which agrees better with theory calculations.  Figure \ref{fig:metac} shows the data and fit for each channel in this analysis. The result from BESIII provides strong evidence that previous inconsistent $\eta_c(1S)$ parameters  in radiative charmonium decays and two-photon collisions$B-$meson decays are caused by the hindered-$M1$ transition and non-resonant interence.

\begin{figure}[!h]
\begin{center}
\includegraphics[height=4 cm]{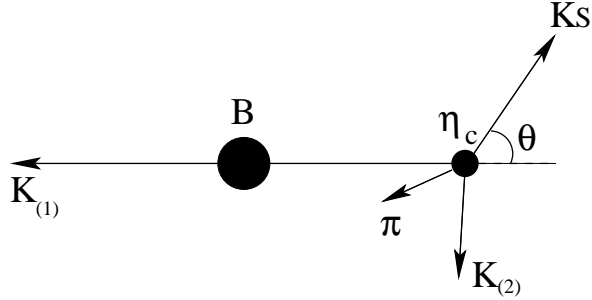}
\caption{The decay $B^{\pm}\to K^{\pm}\eta_c\to K^{\pm}(K_S K\pi)^0$.}
\label{pic:7}
\end{center}
\end{figure}

\begin{figure}[!h]
\begin{center}
\begin{tabular}{ccc}
\includegraphics[height=4 cm]{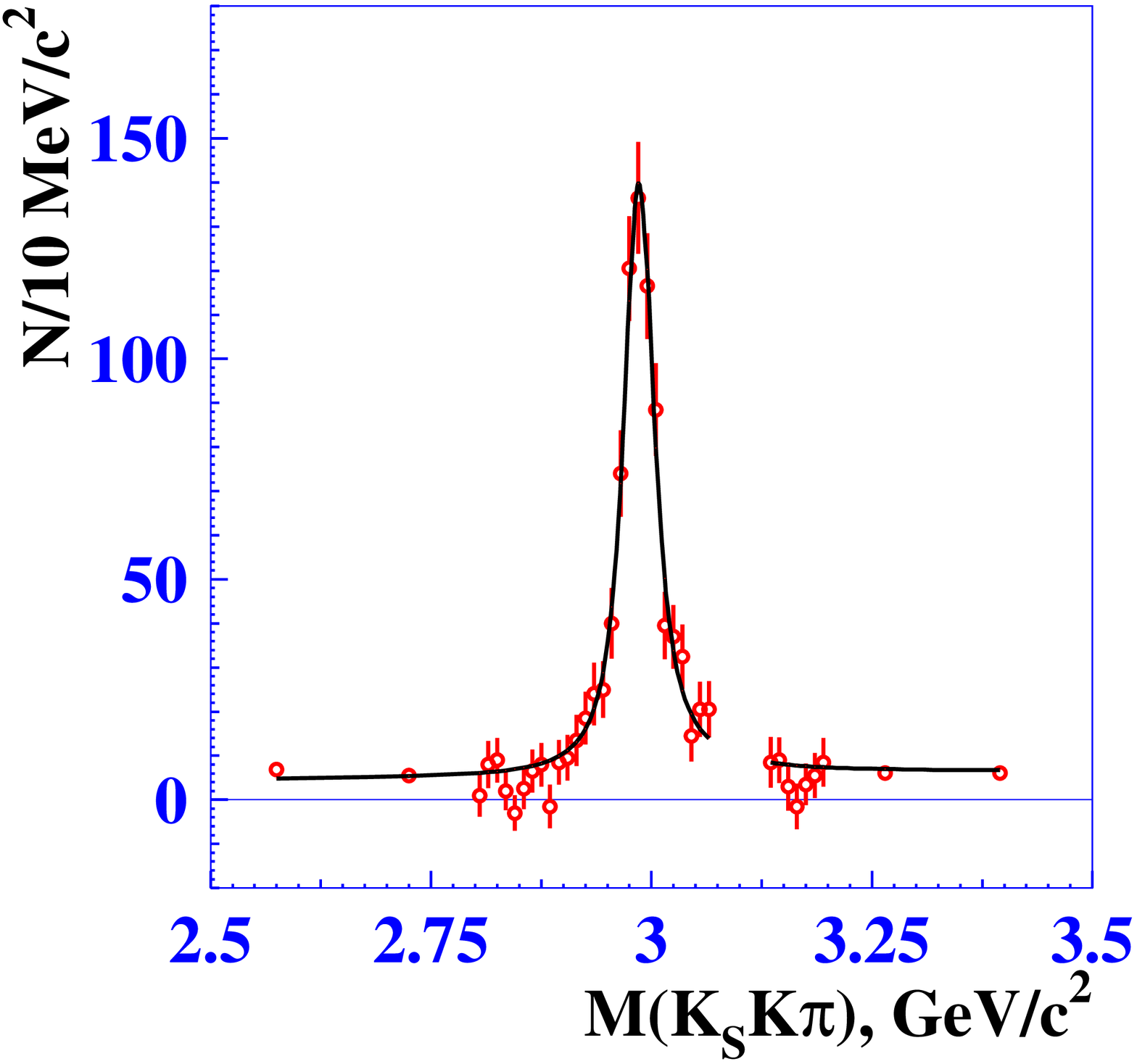}&
\includegraphics[height=4 cm]{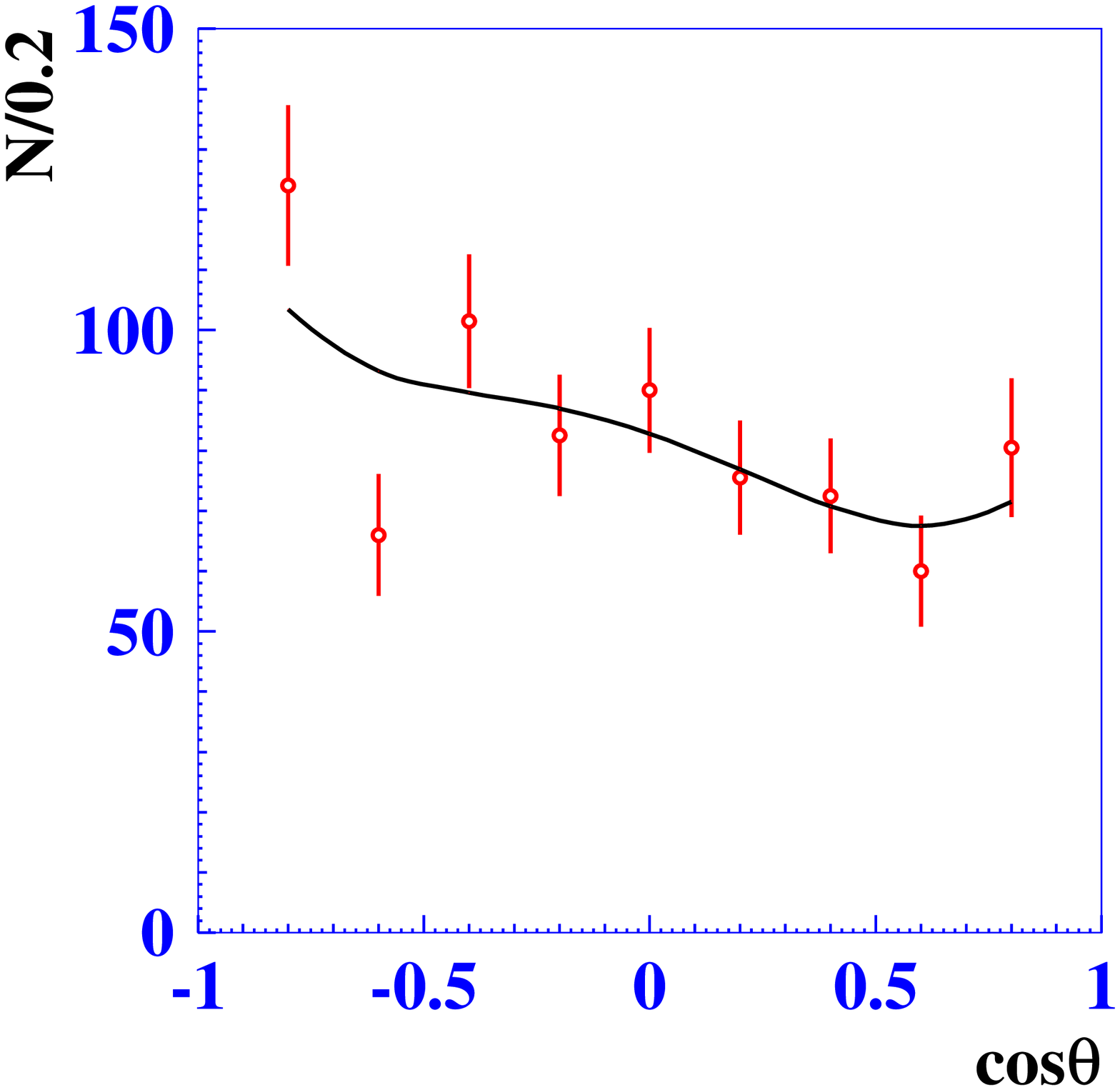}&
\includegraphics[height=4 cm]{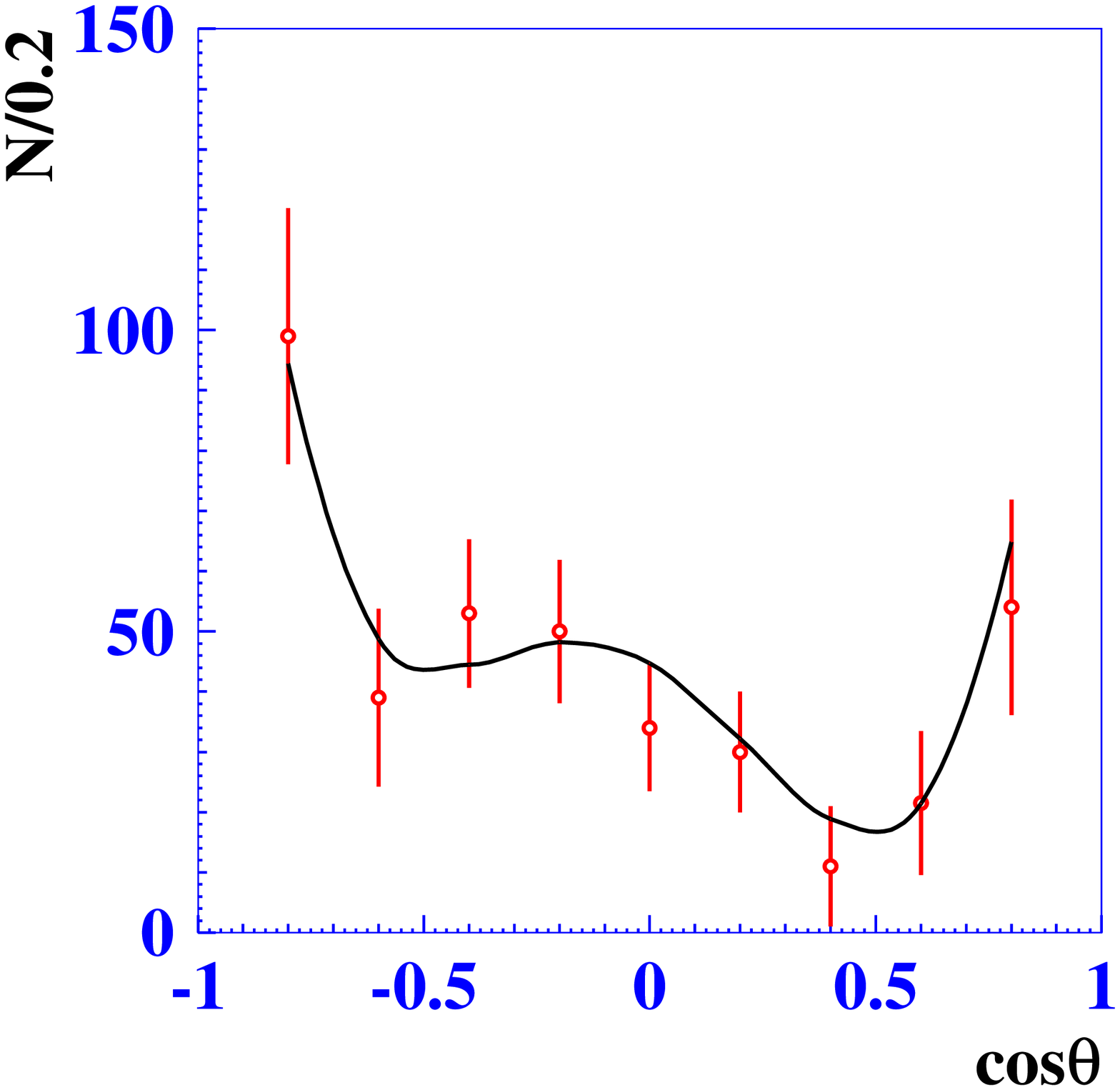}\\
\end{tabular}

\caption{$\eta_c(1S)$ lineshape measurement via $\eta_c\to K_SK\pi$ in $B^{\pm}\to K^{\pm}(K_SK\pi)^0$ from Belle: Projections of the fit in $K_SK\pi$
invariant mass in the $\eta_c(1S)$ mass region (left) and $\cos\theta$ in the $\eta_c(1S)$ invariant 
mass signal (center) and sideband (right) regions.
The combinatorial background
is subtracted. The gap near 3.1 GeV/$c^2$ is due to the $J/\psi$ veto.
The bin size along the
$\cos{\theta}$ axis is $0.2$. Along the $M(K_S K\pi)$ axis the bin size is
$10$ MeV/$c^2$ in
the signal region and $150/130$ MeV/$c^2$ in the left/right sideband 
region}
\label{pic:fit1}
\end{center}
\end{figure}

There are also new measurements of $\eta_c(1S)$ from $B-$factories. With a data sample of 535 million $B\bar{B}$-meson pairs, Belle measured  the $\eta_c(1S)$ lineshape via $B^+\to\eta_c,\eta_c\to K_S K^{+}\pi^{-}$~\cite{Vinokurova:2011dy}. Compared to the two-photon process, the advantages of studying $\eta_c(1S)$ in $B$ decays are the relatively large reconstruction efficiency, small background, and the fixed quantum numbers of the initial state. In Belle's analysis, a 2D-fit to the $M(K_SK\pi)$ and $\cos\theta$ distributions is performed to obtain the interference contributions, where $\cos\theta$ is defined in Fig.~\ref{pic:7}. To reduce the uncertainty from the interference, P- and D-waves are separated from the S-wave in the non-resonant background. Fig.~\ref{pic:fit1} shows projections of the fit. $\eta_c(1S)$ parameters measured in this analysis are $M(\eta_c)=(2985.4\pm 1.5^{+0.5}_{-2.0})$  MeV/$c^2$ and $\Gamma(\eta_c)=(35.1\pm 3.1^{+1.0}_{-1.6})$ MeV/$c^2$, which are consistent with recent results from BESIII~\cite{BESIII:2011ab} and BaBar~\cite{delAmoSanchez:2011bt}.

The $h_c \to \gamma \eta_c(1S)$ transition can provide a new laboratory to study $\eta_c(1S)$
properties. The $\eta_c(1S)$ line shape in the $E1$ transition $h_c\to\gamma\eta_c(1S)$ should
not be as distorted as in two-photon production at $B-$factories and in other charmonium
decays, because non-resonant interfering backgrounds to the dominant transition are small. 
BESIII is trying to use this method to extract $\eta_c(1S)$ parameters~\cite{Olsen:2012xn}.

\section{\bf $\eta_c(2S)$}
The $\eta_c(2S)$ is the first radial excitation of the $\eta_c$ charmonium ground state. It was first observed by the Belle in $B$ decays~\cite{babar_B2Ketacp}. Since then, it has been confirmed and studied in $B-$fatories via two-photon fusion, double-charmonium production and $B$ decays~\cite{babar_B2Ketacp}$-$\cite{belle_Jpsiccbar}. In early days the only known decay mode of $\eta_c(2S)$ was $K_S K^+ \pi^-$~\cite{ref:PDG_2010}.

\begin{figure}[!h]
\begin{center}
\begin{tabular}{ccc}
\includegraphics[height=4 cm]{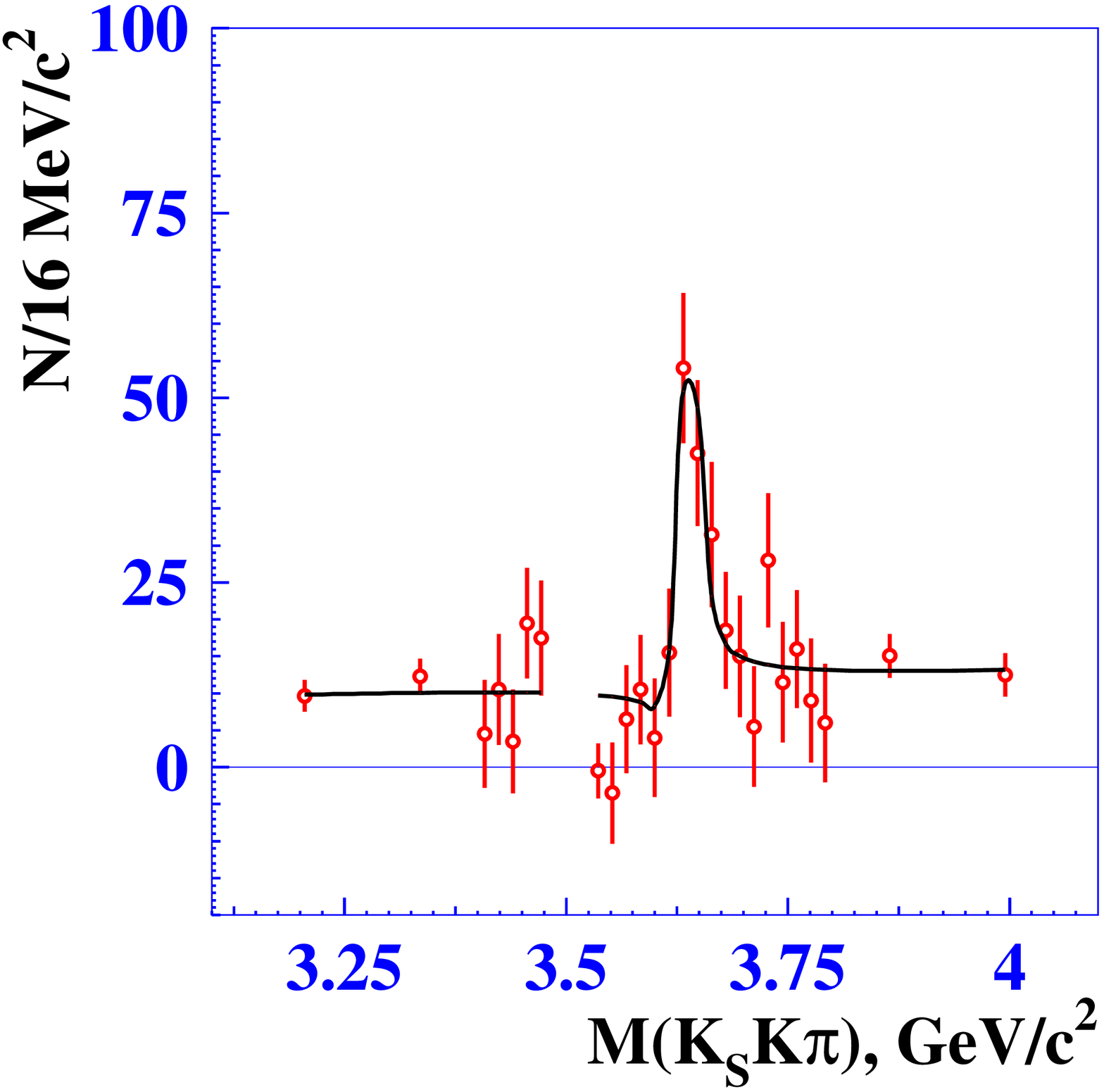}&
\includegraphics[height=4 cm]{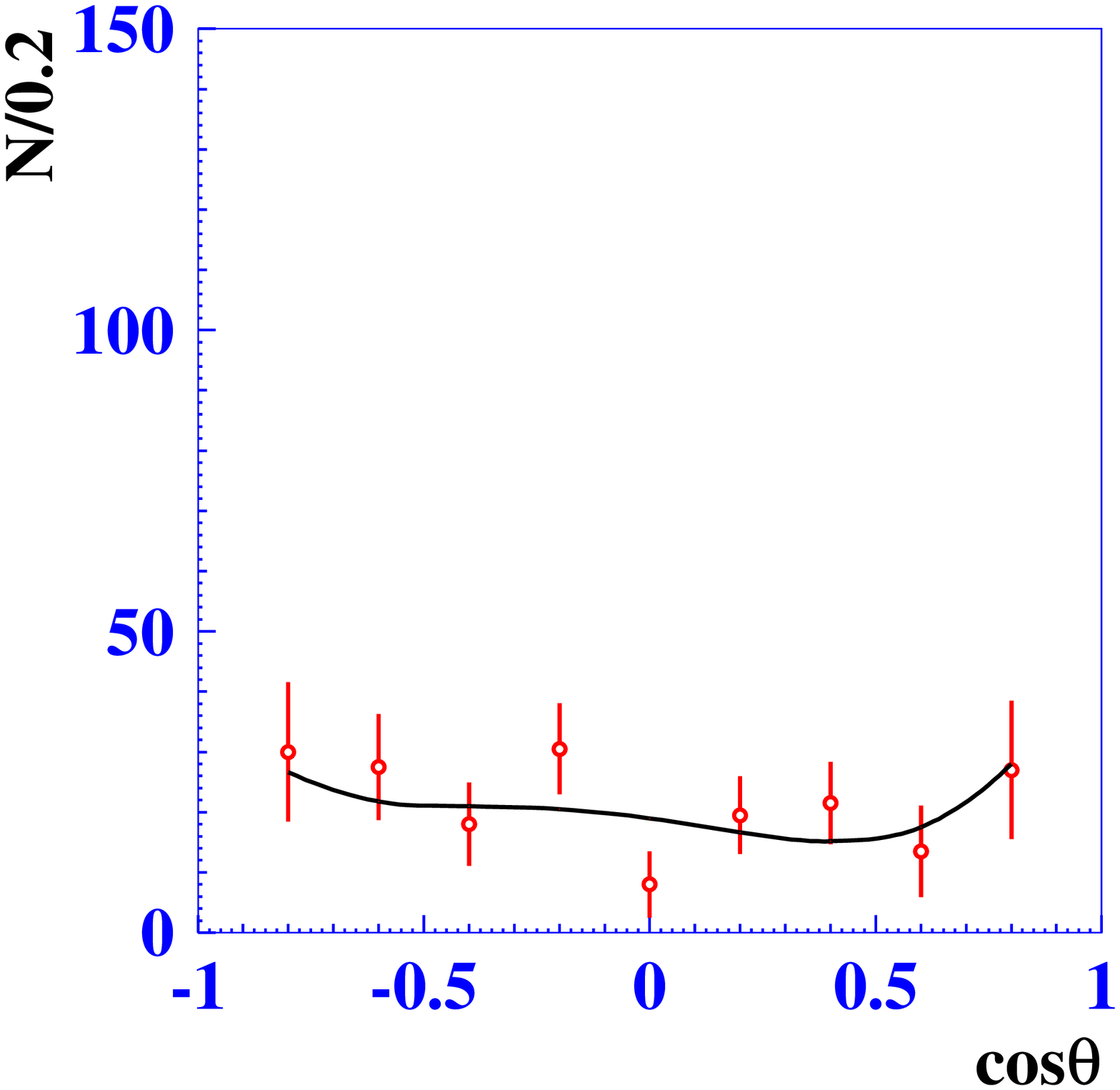}&
\includegraphics[height=4 cm]{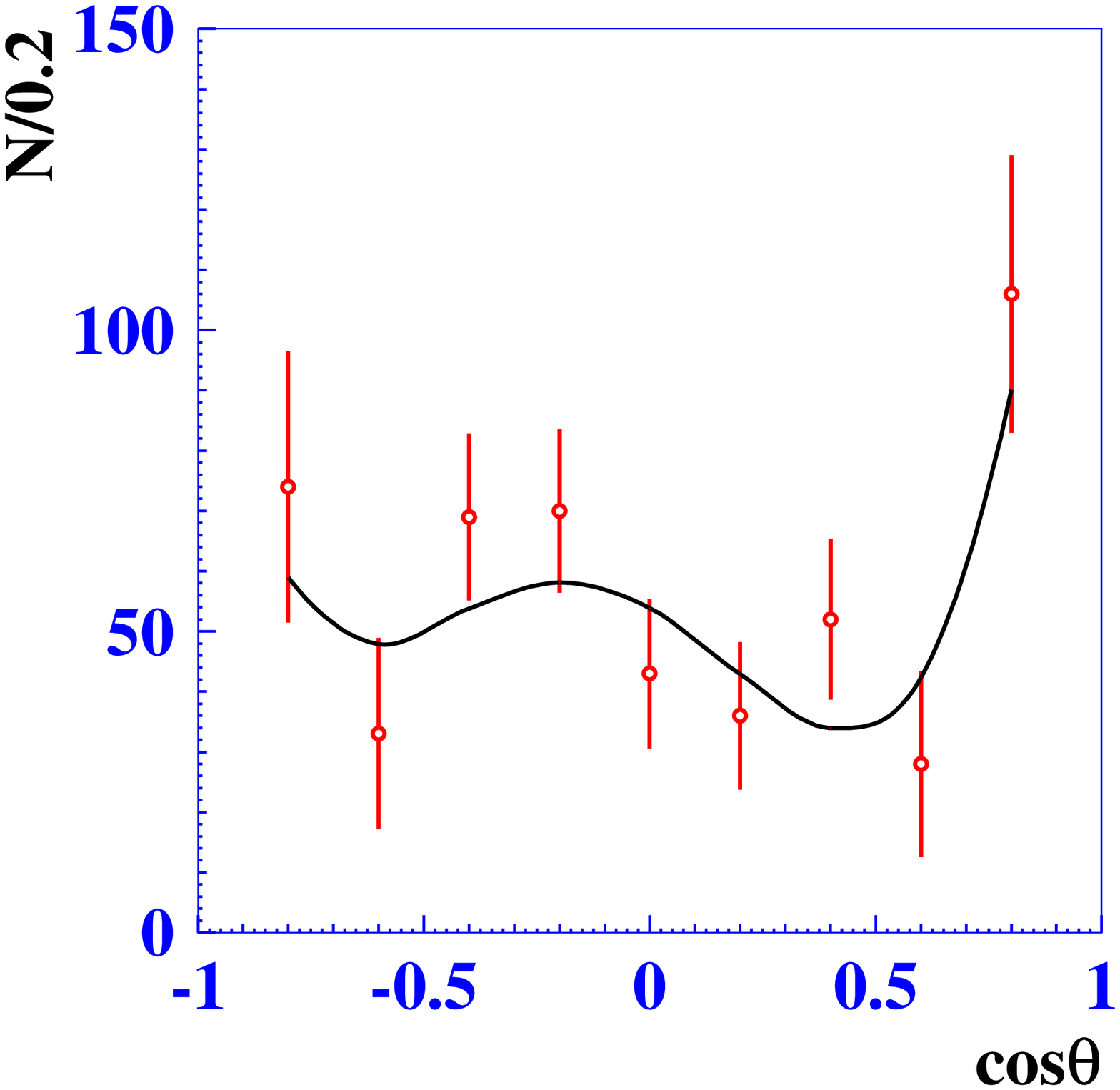}\\
\end{tabular}
\caption{$\eta_c(2S)$ lineshape measurement via $\eta_c\to K_SK\pi$ in $B^{\pm}\to K^{\pm}(K_SK\pi)^0$ from Belle: Projections of the fit in $K_SK\pi$ invariant mass in the $\eta_c(2S)$ mass region (left) and
$\cos\theta$ in the $\eta_c(2S)$ invariant 
mass signal (center) and sideband (right) regions. 
The combinatorial background
is subtracted. The gap near 3.5 GeV/$c^2$ is due to the $\chi_{c1}$ veto.
The bin size along the
$\cos{\theta}$ axis is $0.2$. Along the $M(K_S K\pi)$ axis the bin size is 
$16$ MeV/$c^2$ in the signal region and $130$ MeV/$c^2$ in the sideband region.}
\label{pic:fit2}
\end{center}
\end{figure}

At the present time, Belle's and BaBar's efforts on $\eta_c(2S)$ have been moved to measuring its mass and width and looking for new decays other than $K_S K\pi$. Using the same technique in the study of $\eta_c(1S)$, Belle tried to extract $\eta_c(2S)$ resonance parameters~\cite{Vinokurova:2011dy}. They found that the interference of signal and non-resonant background was very important in the $\eta_c(2S)$ case: with interference,  $M(\eta_c(2S))=3636.1^{+3.9}_{-4.2}(stat.+model)^{+0.7}_{-2.0}$ MeV/$c^2$ and $\Gamma(\eta_c(2S))=6.6^{+8.4}_{-5.1}(stat.+model)^{+2.6}_{-0.9}$MeV/$c^2$; without interference,  $M(\eta_c(2S))=3646.5\pm 3.7^{+1.2}_{-2.9}$ MeV/$c^2$ and $\Gamma(\eta_c(2S))=41.1\pm 12.0 ^{+6.4}_{-10.9}$ MeV/$c^2$. Figure \ref{pic:fit2} shows projections of the fit in Belle's analysis with the consideration of interference.

 \begin{figure}[!h]
   \begin{center}
     \includegraphics[scale=0.48]{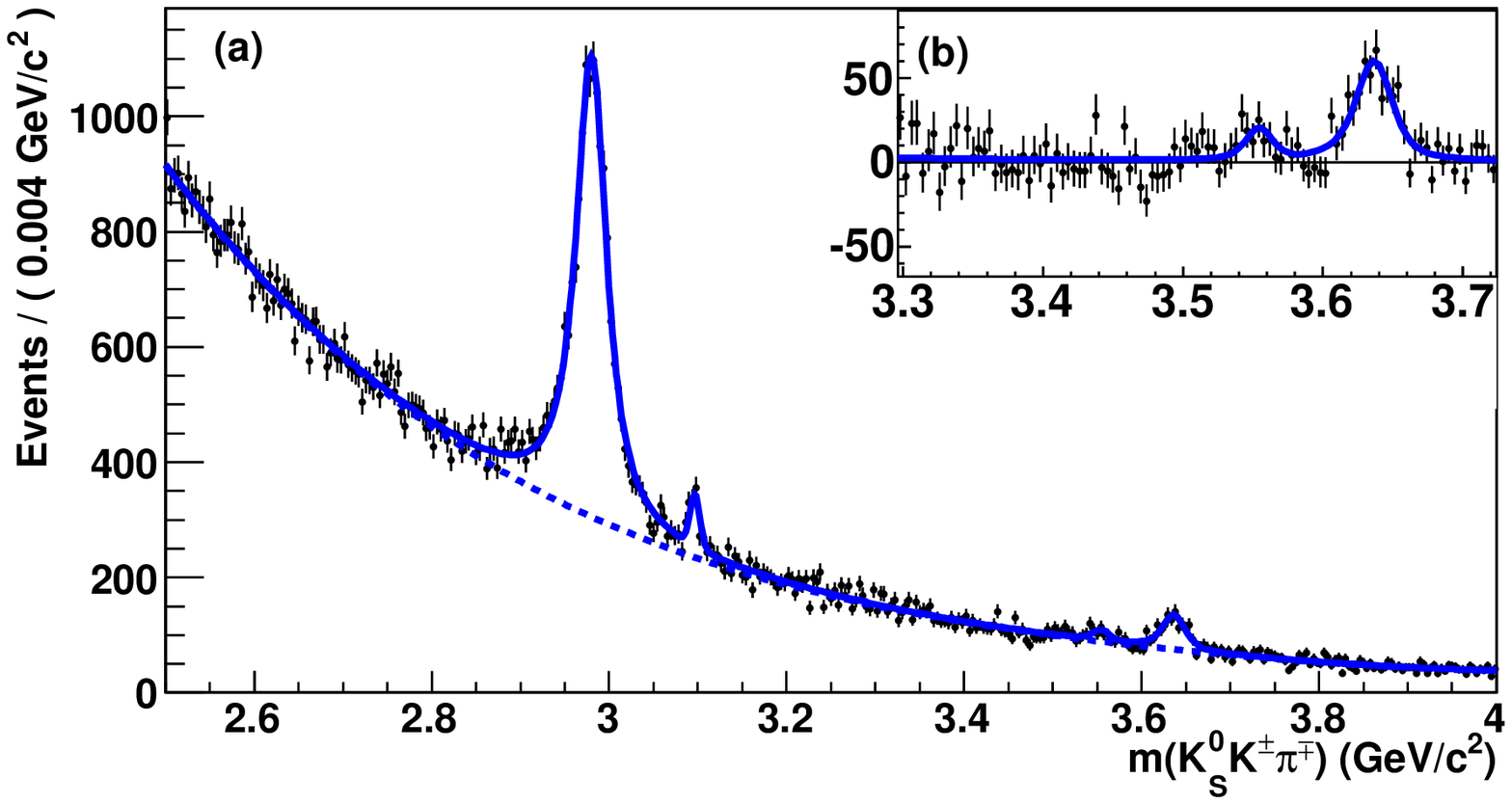}
     \includegraphics[scale=0.48]{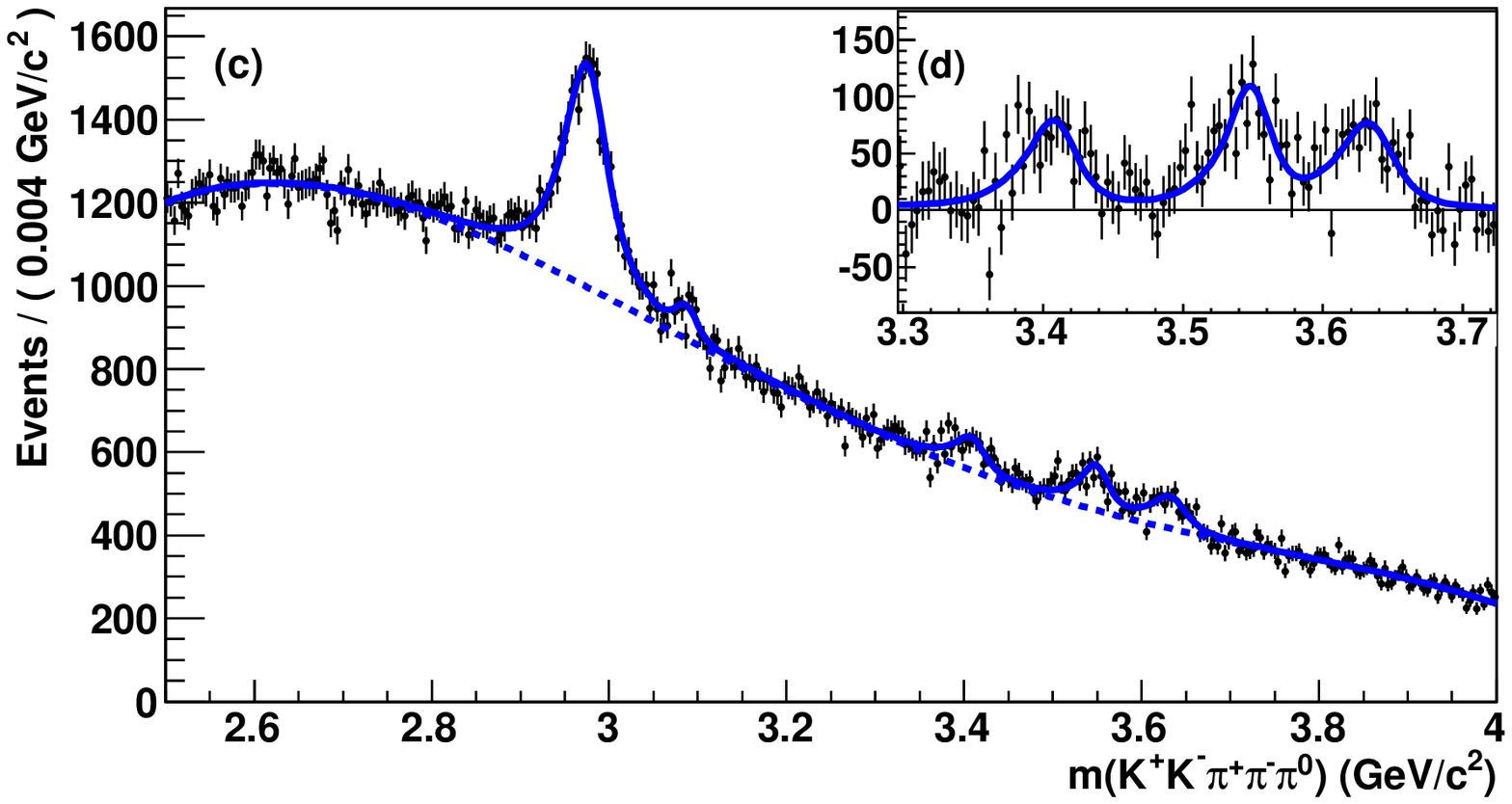}
     \caption{$\eta_c(1S)$ and $\eta_c(2S)$ decays to  the $K_S K^+ \pi^-$\ and the $K^+ K^- \pi^+\pi^-\pi^0$ in two-photon interactions from Babar: Fit to (a) the $K_S K^+ \pi^-$\ and (c) the $K^+ K^- \pi^+\pi^-\pi^0$ mass spectrum. The
       solid curves represent the total fit functions and the dashed curves
       show the combinatorial background contributions. The
       background-subtracted distributions are shown in (b) and (d),
       where the 
       solid curves indicate the signal components. 
     }
     \label{fig:fit}
   \end{center}
 \end{figure}

BaBar studied $\eta_c(1S)$ and $\eta_c(2S)$ in the two-photon processes $\gamma\gamma\to K_S K^+\pi^-$ and $\gamma\gamma\to K^+K^-\pi^+\pi^-\pi^0$ using a data sample of 519.2~fb$^{-1}$ near the $\Upsilon(nS)$ ($n = 2,3,4$) resonances. $\eta_c(2S)\to K^+K^-\pi^+\pi^-\pi^0$ was found with a significance of  $5.3\sigma$~\cite{delAmoSanchez:2011bt}. BaBar also obtained the $\eta_c(2S)$ mass and width by performing to a fit to the $M(K_S K^+\pi^-)$ spectrum (Fig.~\ref{fig:fit}), and obtained the values $M(\eta_c(2S))=3638.5 \pm 1.5 \pm 0.8$~MeV/$c^2$ and $\Gamma(\eta_c(2S)) = 13.4\pm4.6\pm 3.2$~MeV/$c^2$.

\begin{figure*}[htb]
\centering
  \includegraphics[width=0.49\textwidth]{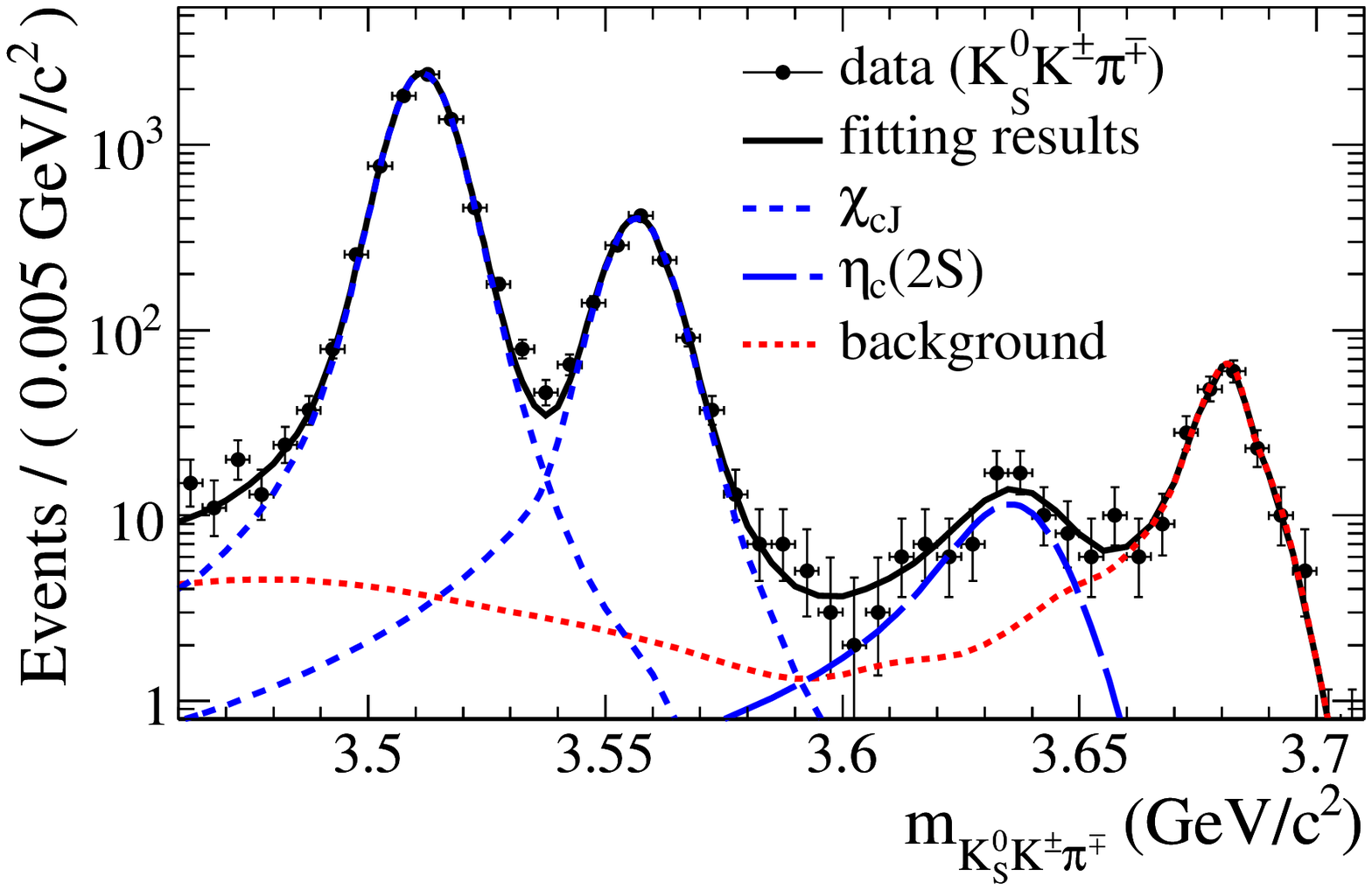}
  \includegraphics[width=0.49\textwidth]{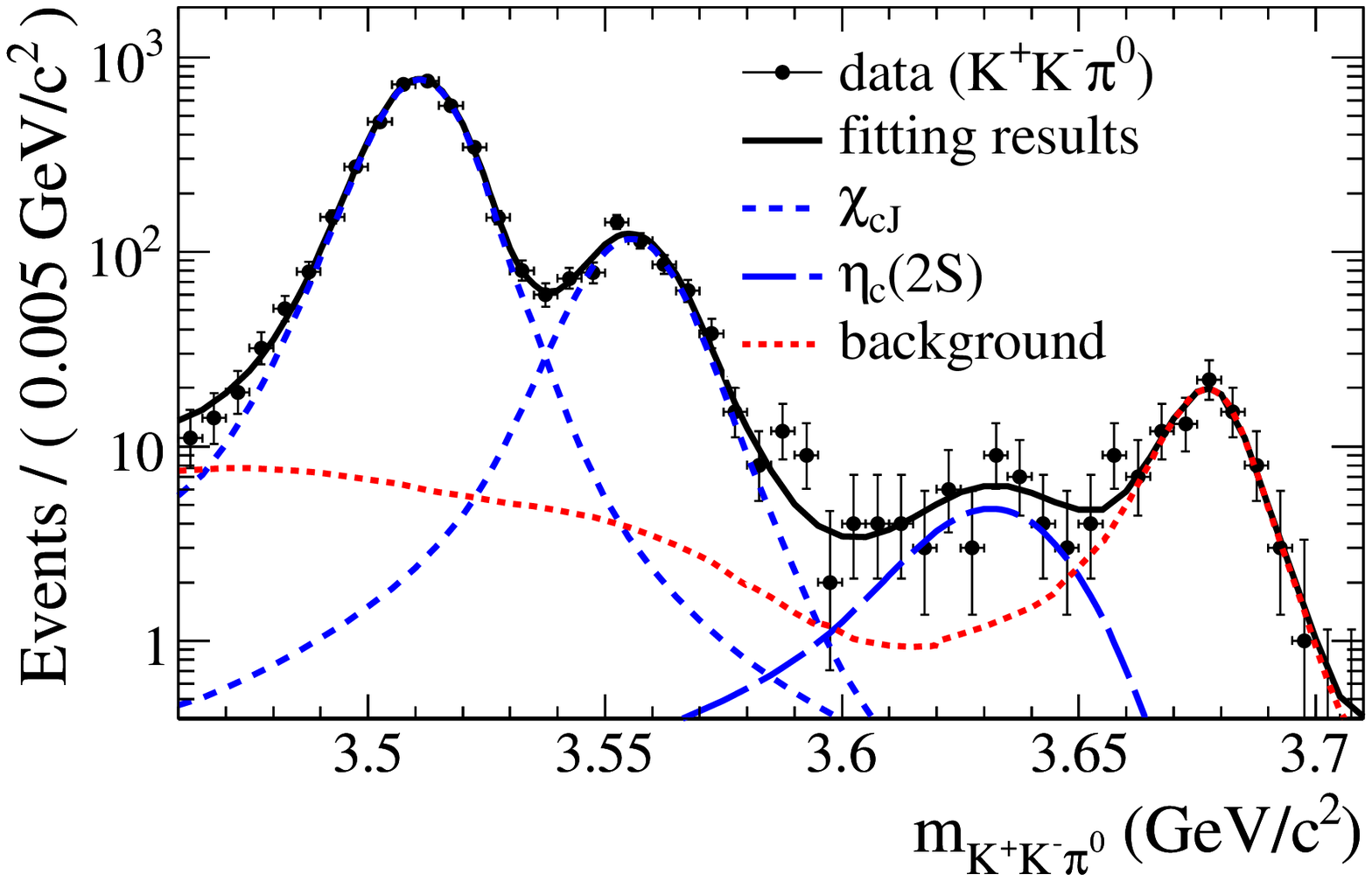}
\caption{Observation of $\eta_c(2S)$ in $\psi'\to\gamma KK\pi$  from BESIII: The invariant-mass spectrum for $K_S K^+ \pi^-$ (left panel),
$K^+K^-\pi^0$ (right panel), and the fit to the data.} \label{pic_fit_etacp}
\end{figure*}

On the other hand, the search for $\eta_c(2S)$ through a radiative transition from the
$\psi'$ is very hard and has been stuck for many years. The difficulty comes from the detection of the low-energy radiative photon in the $\psi'\to\gamma\eta_c(2S)$. The branching ratio and mechanism of this process has been predicted by many papers, but the absence of experimental result had made it impossible to discriminate among them~\cite{gaoky}$-$\cite{mabq}. For a long period, this transition has been looked for by Crystal Ball~\cite{cbal}, BES~\cite{ycz}, CLEO~\cite{etacp_cleo_c} and BESIII~\cite{etacp_VV}, but none of them could provide a convincing observation. Most recently, BESIII made the first
observation of this process using the 106M $\psi'$ sample. Analyses of $\psi'\to
\gamma\eta_c(2S)$ with $\eta_c(2S)\to K_S K^+\pi^-$ and $K^+K-\pi^0$ gave a significance greater than $10\sigma$. 
In addition to the excellent low energy photon detection, smart use of the kinematic fitting plays a key role in this observation.
Fig.~\ref{pic_fit_etacp} shows data and fits. Numerical results are $M(\eta_c)=3637.6\pm 2.9\pm 1.6$~MeV/$c^2$), 
$\Gamma(\eta_c(2S))=16.9\pm 6.4\pm4.8$~MeV  and $B(\psi'\to \gamma\eta_c(2S))\times B(\eta_c(2S)\to
K\bar K\pi) = (1.30\pm 0.20\pm 0.30)\times 10^{-5}$. The branching ratio of $\psi'\to\gamma\eta_c(2S)$ is determined to be $B(\psi'\to\gamma\eta_c(2S)) = (6.8\pm 1.1\pm 4.5)\times 10^{-4}$. These results are consistent with results from $B-$factories.

\newpage
\section{\bf Summary}

In summary, experimental studies of $h_c$, $\eta_c(1S)$ and $\eta_c(2S)$ have made major progress and still face nontrivial challenges:

\begin{itemize} 

\item $h_c$

The key branching ratios $\psi'\to\pi^0 h_c$ and $h_c\to\gamma\eta_c$ have been nailed down, so the absolute $h_c$ cross sections/branching ratios are available. A new prolific production mode of $h_c$ has been found: $e^+e^-\to\pi^+\pi^- h_c$. Because the $B(h_c\to\gamma\eta_c)$ is about $50\%$, the remaining decays of $h_c$ should be large enough to be observed. Further measurement of these unclear decays will be helpful to understand the property of $h_c$ and the transition mechanism between $h_c$ and other charmonium.

\item $\eta_c(1S)$

The mass and width are more consistent in $\psi'$  decays, $B$ decays and $\gamma\gamma$ production than previously, and the charmonium $1S$ mass hyperfine splitting  from experiments agrees better with theory. The $\eta_c$ lineshape in $h_c$ is not as distorted as in charmonium/$B$ decays and  $\gamma\gamma$ fusion, because of the small non-resonant interfering background. Ultimately, with a large $\psi'$ sample, this channel will be best suited to determine $\eta_c$ resonance parameters.

\item $\eta_c(2S)$

Finally, $\eta_c(2S)$ was observed in charmonium transitions after thirty years of searching. Finding $\eta_c(2S)$ has proved hard enough, but understanding its decay properties and measuring its lineshape are proving more difficult. Due to the lower production rate, $\eta_c(2S)$ is hard to reconstruct and is affected by interference more than $\eta_c(1S)$. Further measurement of $\eta_c(2S)$ will require large statistics, a convincing theoretical model and a sophisticated fitting method.

\end{itemize}

\end{document}